\documentstyle[prb,aps,twocolumn]{revtex}
\begin{document}
\draft
\title{Electronic structure of quasi free 2 dimensional HfS$_2$ clusters}
\author{M.~Traving, C.~Kreis, R.~Adelung, L.~Kipp$^*$, and M.~Skibowski}
\address{Institut f\"ur Experimentelle und Angewandte Physik, Universit\"at Kiel,
D--24118 Kiel, Germany}
 
\date{\today}
\maketitle
\begin{abstract}

Cluster research so far only dealt with either 3D free clusters or clusters
on substrates with strong electronic coupling at the interface. Preparation
of free and orientated 2D clusters is usually not possible because one of
these conditions generally excludes the other. We demonstrate a new way to
grow 2D clusters of transition metal dichalcogenides which are only weakly
 bound to the substrate, i.e. quasi
free, but all show the same orientation thus allowing angle resolved
photoemission experiments. In combination with scanning tunneling microscopy
this provides for the first time the possibility to investigate the
correlation between the size of quasi free 2D clusters and its associated
electronic structure. For 2D $\rm HfS_2$ clusters grown by van der Waals epitaxy we
demonstrate that k$_\parallel$--dispersion of electronic states 
         already occurs for  cluster diameters above $\sim$100~nm.
\end{abstract}

In the last decade there was a growing interest in the geometric and
electronic properties of clusters. Especially, fundamental properties as
band gaps or melting points depend strongly on the cluster size
\cite{Alivis96,Brus91,VoKaGi94}. According to Alivisatos \cite{Alivis96} in
the course of growing cluster size first the formation of the centers of the
electronic bands can be observed which coincide with the corresponding
atomic levels and afterwards the development of the band edges. 
With further increasing cluster size the investigation of the formation of the
momentum-resolved electronic structure of intrinsic clusters is challenged. 
Employing angle resolved photoemission spectroscopy two
conditions need to be satisfied: (i) The clusters must be quasi free and
 independent of the substrate to retain their
intrinsic properties and (ii) because of spatial averaging they have to 
have the same orientation to allow momentum-resolved measurements. 
In general one of these conditions excludes the other and vice versa. 
So far
free clusters could only be prepared in three or zero dimensions. Flying
''disks'' (2D) or ''wires'' (1D) have not been reported. Oriented clusters of such
dimension have been prepared on surfaces of a number of preferably
covalently bonded bulk crystalline materials (see e.g.\ ref.\
\onlinecite{Brus91}). However, strong bonding between cluster and substrate
together with additional dangling bond surface states on the cluster surface
complicate the investigation of intrinsic properties of the cluster itself.

In this paper we utilize the extremely weak van der Waals forces 
between sandwiches of quasi two dimensional layered transition metal
 dichalcogenides (TMDC) to prepare quasi free 2D HfS$_2$ clusters.
  Extremely flat WSe$_2$ surfaces serve as ideal substrates to study 
  intrinsic properties of the clusters independent of the underlying
   material. Crystal structures of layered TMDC's are characterized by
    hexagonal metal layers sandwiched between chalcogen sheets.  While 
    inside one sandwich layer strong ionic and covalent bonding predominates, 
    weak van der Waals forces act between the layers \cite{WilYof69,Liang86,LieTer77}
     allowing heteroepitaxy of layered materials almost independently of the actual
      lattice mismatch \cite{KoSuI84,KoSuII84,Koma92,OhPaUe90,ScLoLa95,TiSePe94}. 
      In addition, no dangling bond type surface states have been observed on 
      these surfaces. The weak interaction between the sandwich layers is 
      responsible for the small dispersion of the electronic states in 
      normal direction of layered materials \cite{KlLaSc98}.

Van der Waals epitaxy (vdWE) is applied here to grow HfS$_2$ clusters
 on WSe$_2$. Employing angle resolved photoemission spectroscopy  in combination
  with scanning tunneling microscopy we focus on the development  of the
   dispersion E(k$_\parallel$) of  electronic states parallel to the surface
    with increasing cluster size.\\ 

All experiments were carried out in an ultrahigh vacuum (UHV) system (base
 pressure 10$^{-10}$~mbar) consisting of chambers for (i) van der Waals 
 epitaxy (vdWE), (ii)  angle resolved photoemission spectroscopy (ARPES), 
 (iii) scanning tunneling microscopy (STM), and (iv) low energy electron
  diffraction (LEED), all connected by an UHV transfer system.

HfS$_2$ clusters were grown by coevaporation of hafnium and sulfur from 
 different sources. The hafnium beam was generated by an electron beam
  evaporator \cite{Omicron} with an integrated flux monitor permitting 
  direct control of the hafnium flux during epitaxy. Sulfur was evaporated
   from a Knudsen cell by heating pyrite (FeS$_2$) to about 400$^{\circ}$~C.
    The corresponding partial pressure of sulfur was 
    $\rm p_{S_2}=1\times10^{-9}$~mbar. Substrate temperatures T$_{S}$
     were chosen between 300 and 400$^{\circ}$~C. The 2H--WSe$_2$
      substrates and 1T--$\rm HfS_2$ crystals used as substrate and
       reference material, respectively, were grown by chemical
        vapor transport using iodine as transport agent. All 
        substrates were air cleaved and annealed at 450$^{\circ}$~C 
        for two hours to obtain smooth and clean surfaces.  

Angle resolved photoemission spectra were taken with a Helium discharge 
lamp (HeI$_{\alpha} radiation, h\nu=21.22$~eV). Electrons were detected
 by a $180^\circ$ spherical analyzer (energy resolution $\rm \Delta E=65$~meV)
  mounted on a two axes goniometer 
  (angle resolution $\Delta \vartheta < 0.25^\circ$). The STM
   \cite{Omicron} consists of a single tube scanner system  mounted
    on a self developed vibration damping with an eddy current attenuation.
     Coverages were determined by counting the pixels in STM images that
      belong to either  HfS$_2$ clusters or the substrate.\\

In fig.\ \ref{HfS2} we show a series of combined normal emission photoelectron
 and scanning tunneling microscopy data in different phases of $\rm HfS_2$ growth
  on bulk $\rm WSe_2$. (The middle panel shows difference spectra obtained by 
  subtracting a normalized clean $\rm WSe_2$ spectrum from data shown in the 
  left panel. For comparison the differences are hatched in the left panel.)
   At the bottom a normal emission photoelectron spectrum of the cleaved 
   $\rm WSe_2$ substrate exhibiting the characteristic valence band emission
    peaks \cite{TrBoKi97} and the corresponding STM image and LEED pattern are
    plotted. Upon initial stages of growth the STM image for a coverage of 52\%
     of a monolayer (ML) $\rm HfS_2$ reveals small irregularly shaped clusters with
      typical diameters of about 50~nm. The corresponding photoemission spectrum 
      is still being mainly attributable to emissions of the $\rm WSe_2$ substrate.
       With increasing cluster size up to a coverage of 87\% the upper valence band
        peak A corresponding to $\rm WSe_2$ is still visible while peak B is covered
         by a new structure which can be related to the $\rm HfS_2$ emission feature
          R (compare top panel). At 92\% coverage with respect to the first monolayer
           peak A still appears as a weak shoulder. For coverages beyond one 
           monolayer (STM image at the top) the photoemission spectrum reveals 
           the significant structures of pure $\rm HfS_2$. For comparison see
            the spectrum of a cleaved $\rm HfS_2$ crystal shown at the top. In 
            the spectra the peaks of the WSe$_2$ substrate are broadened in the
             course of the epitaxial growth \cite{Horn90}. Especially, WSe$_2$
              emission A shows this behavior.

Just like the photoemission spectrum at 80\% coverage in the first monolayer
 the corresonding LEED image consists of a composition of both $\rm HfS_2$ 
 and $\rm WSe_2$ contributions: Two hexagonal patterns are superimposed with
  the outer pattern belonging to $\rm WSe_2$ because of its smaller lattice
   parameter (${\rm a_{WSe_2}\,=\,3.286~\AA}$ \cite{ManSki94}, $\rm{a_{HfS_2}\,=\,3.635~\AA}$~\cite{TagWad58}). It should be noted that 
the 2D clusters grow with the same lattice parameters parallel to the
 surface as bulk HfS$_2$. The alignment of the HfS$_2$ with the WSe$_2$
  pattern demonstrates that the HfS$_2$ clusters have the same 
  crystallographic orientation as the substrate lattice. This has
   also been observed for several other heterojunctions (e.g.~GaSe
    growth on WSe$_2$ \cite{LaScI94} and InSe(GaSe) growth on GaSe(InSe)
     \cite{LaKlSc95}). The alignment offers the possibility to examine
      the k$_{\parallel}$--dispersion of the electronic structure of
       quasi free 2D HfS$_2$ clusters as will be discussed later.

Note that in the submonolayer regime the photoelectron spectra of the HfS$_2$
 clusters are super\-imposed by WSe$_2$ substrate contributions and, especially,
  at small coverages are hardly noticable. In order to examine the electronic
   structure of the HfS$_2$ clusters in more detail, the HfS$_2$ overlayer spectrum
    has to be separated from the substrate features. Since the HfS$_2$ clusters 
    are only bound by weak van der Waals forces they may be viewed as nearly 
    independent of the substrate. This justifies substractions of the pure WSe$_2$ 
    spectrum to separate spectral contributions of clusters and substrate.

WSe$_2$ emission A ($\vartheta$~=~0$^\circ$) is located in the
 band gap of HfS$_2$ and is only weakened by the HfS$_2$ clusters but not
 superimposed by HfS$_2$ emissions. Consequently, it remains discernable 
 up to coverages of more than one monolayer. This provides the possibility 
 to estimate the contribution of the WSe$_2$ substrate to the photoelectron
  spectra of both, clusters and substrates. For each coverage we compared the
   area under peak A before and after deposition of the HfS$_2$ clusters and
    thus considered relative height and  broadening  of the substrate spectrum 
    to obtain a normalized WSe$_2$ spectrum. Subtracting this normalized
     spectrum from the combined HfS$_2$/WSe$_2$ spectra results in the
      difference spectra shown in fig.~\ref{HfS2} (middle panel). 
This analysis yields further details on the electronic structure of the
 clusters. While the HfS$_2$/WSe$_2$ spectrum at 72\% coverage is
  dominated by substrate contributions, the difference spectrum
   already reveals the weak contributions of the HfS$_2$ clusters. 
   For higher coverages the difference spectrum resembles the HfS$_2$
    bulk spectrum more clearly than the HfS$_2$/WSe$_2$ spectrum.

Following the combined characterization of the geometric and electronic
 features of the HfS$_2$ clusters we will now investigate the electronic
  structure of the clusters parallel to the surface. Fig.~\ref{winkel} 
  (a) and (b) show the change of the photoemission spectra with increasing
   cluster sizes and final formation of a complete monolayer at two off
    normal emission angles  ($\vartheta$~=~20$^\circ$, $\vartheta$~=~38$^\circ$)
     along the $\Gamma$K(AH) direction of the Brillouin zone (left panels). 
     As in fig.\ref{HfS2} the corresponding difference spectra are plotted 
     in the right panels.

In line with the results in normal direction at 72\% coverage 
(650~K substrate temperature) corresponding to cluster diameters
 of appoximately 100~nm (see fig.~\ref{HfS2}) new structural 
 features appear which in the case of $\vartheta$~=~38$^\circ$
  can be attributed to the HfS$_2$ emission feature S. This is
   further corroborated for cluster formation at 600~K substrate 
   temperature. At a coverage of 80\% typical cluster sizes are 
   comparable to those observed for 72\% coverage at 650~K while
    shapes of the latter are smoother because of the higher substrate
     temperature. The photoemission ratio clusters to substrate 
     is higher in this case because of the higher coverage and 
     spectral features corresponding to the clusters can be
      seperated more clearly. At 80\% coverage the characteristic
       emission features of HfS$_2$ R,S,T,U, and V emerge and a
        distinct change in the density of states for the three polar
         angles $\vartheta$ is observable (see fig.~\ref{winkel}).
          Fig.~\ref{dampfpes}(a) shows the photoemission spectra of
           cleaved bulk HfS$_2$ in $\Gamma$K(AH) direction where the
            spectra for $\vartheta$~=~0$^\circ$, $\vartheta$~=~20$^\circ$,
             and $\vartheta$~=~38$^\circ$ are replaced by the difference 
             spectra of van der Waals epitaxially grown HfS$_2$ at 80\% 
             coverage. They qualitatively well describe the band 
             dispersion along $\Gamma$K(AH). This demonstrates that
              the main HfS$_2$ spectral features are developed revealing 
              k$_\parallel$--dispersion of the electronic states even 
              for cluster diameters of about 100~nm.

Although the photoelectron spectra of the van der Waals epitaxially 
grown films resemble those of bulk HfS$_2$ well, slight but intriguing
 differences exist. At $\vartheta$~=~38$^\circ$ especially emission S 
 is shifted towards lower binding energies demonstrating less pronounced
  k$_\parallel$--dispersion of the bandstructure in clusters as compared
   to bulk HfS$_2$. Further differences between epitaxially grown and 
   cleaved bulk HfS$_2$ become obvious by comparing relative peak positions 
   in normal emission photoelectron spectra. Fig.~\ref{dampfpes}(b) shows
    photoemission spectra of bulk HfS$_2$ and 3~monolayers HfS$_2$ on 
    WSe$_2$. The epitaxial HfS$_2$ emission features are broadened and,
     in particular, a distinct deviation in the relative energy position
      of the peaks occurs (fig.~\ref{dampfpes}(b)). 
      Peak R of 3~monolayers HfS$_2$ shows almost the same 
      binding energy as the corresponding peak R of bulk HfS$_2$. 
      However, emissions U and T of the epilayer spectrum are
       shifted by 350~meV towards lower binding energies with
        respect to the bulk crystal.\\
This modified electronic structure may be caused by the spatial 
restriction of the wave function of the 2D HfS$_2$ clusters. In 
parallel direction the wave functions are limited by the cluster
 boundaries. Dangling bonds at these boundaries may lead to additional
  edge derived electronic states. Perpendicular to the layers the 
  thickness of 3~monolayers is still insufficient to build up a 
  dispersive k$_\perp$ band structure.

In conclusion, applying van der Waals epitaxial growth of layered 
transition metal dichalcogenides we prepared quasi free two dimensional
 clusters of HfS$_2$ on WSe$_2$ substrates. Weak van der Waals bonding
  between cluster and substrate allowed an investigation of the dispersion
   of electronic states with increasing cluster size independent of the 
   substrate. At initial stages of growth the clusters are irregularly 
   shaped rounding off with increasing cluster size and finally forming
    complete monolayers. Due to orientational order of the clusters in 
    the submonolayer regime angle resolved photoemission experiments were
     possible. The spectra exhibit an onset of dispersion of electronic
      states already for typical cluster sizes of about 100~nm as
       characterized by STM. This highlights the necessity of
        orientating quasi free two dimensional clusters on van der 
        Waals surfaces to properly perform k-resolved measurements 
        of the electronic structure of two dimensional clusters 
        as far as combined angle and spatial resolving instruments
        are not available.\\

This work was supported by
the BMBF (project no. 05 SE8 FKA).

  \begin{figure}[t]
    \unitlength1mm
    \caption[HfS2]{\label{HfS2} Photoemission data (h$\nu$=21.22~eV,
     normal emission, k$_{\parallel}=0$) (left panel), difference spectra
      obtained by subtracting a normalized clean $\rm WSe_2$ spectrum 
      (middle panel), and scanning tunneling microscopy data with LEED images
       as inset (E$_{kin}=$100--150~eV) (right panel) showing the transition 
       from the clean $\rm WSe_2$ substrate to the grown $\rm HfS_2$ overlayer.
        Bottom: Photoemission spectrum of the substrate (cleaved and annealed
         $\rm WSe_2$) and corresponding STM image (horizontal size 9~nm). Upward 
         increasing coverage with $\rm HfS_2$: photoemission spectra and STM 
         images of the same samples (horizontal size 150~nm). Coverages are 
         given with respect to one nominal monolayer (100\%). Top: spectrum
          of cleaved $\rm HfS_2$ for reference. Substrate temperatures were
           650~K (52\%,72\%) and 600~K (80\%--3ML), respectively.}
  \end{figure}

\begin{figure}[ht]
    \unitlength1mm
    \caption[winkel]{\label{winkel}
Photoelectron spectra of HfS$_2$/WSe$_2$ for increasing coverage at two 
different polar angles $\vartheta$ in $\Gamma$K(AH) direction ((a)
 $\vartheta$~=~20$^\circ$, (b) $\vartheta$~=~38$^\circ$, left panel
  respectively) in comparison to the corresponding spectra of cleaved 
  HfS$_2$ on top (h$\nu$~=~21.22~eV, $\Delta$E~=~65~meV). (a) and (b),
   right panels: difference spectra obtained by subtracting a normalized
    WSe$_2$ spectrum.}
  \end{figure}

\begin{figure}[ht]
    \unitlength1mm
    \caption[dampfpes]{\label{dampfpes}
(a) Photoelectron spectra (h$\nu$~=~21.22~eV, $\Delta$E~=~65~meV)
 in $\Gamma$K(AH) direction of cleaved bulk HfS$_2$ containing the
  difference spectra of quasi free 2D free HfS$_2$ clusters on
   WSe$_2$ (bold dots, marked by arrows). Typical cluster diameters 
   are $\sim$100~nm (80~$\%$ coverage) (b) photoemission spectra of
    cleaved bulk HfS$_2$ (top) and 3 monolayers HfS$_2$ on WSe$_2$ 
    (bottom): relative peak positions are compared.                     
            }
  \end{figure}


\begin{thebibliography}{10}
\bibitem[*]{}
 Author to whom correspondence should be addressed.
\end{thebibliography}

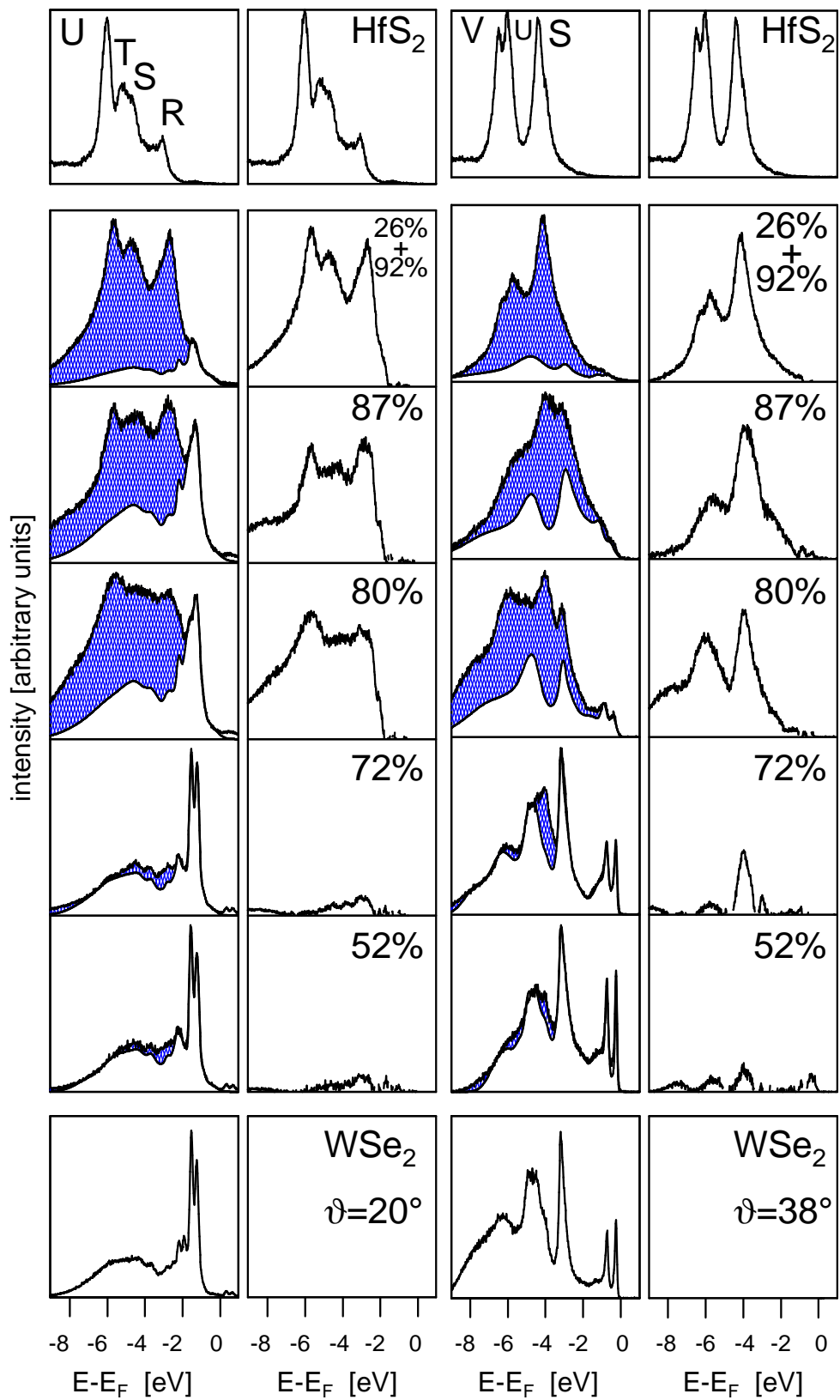
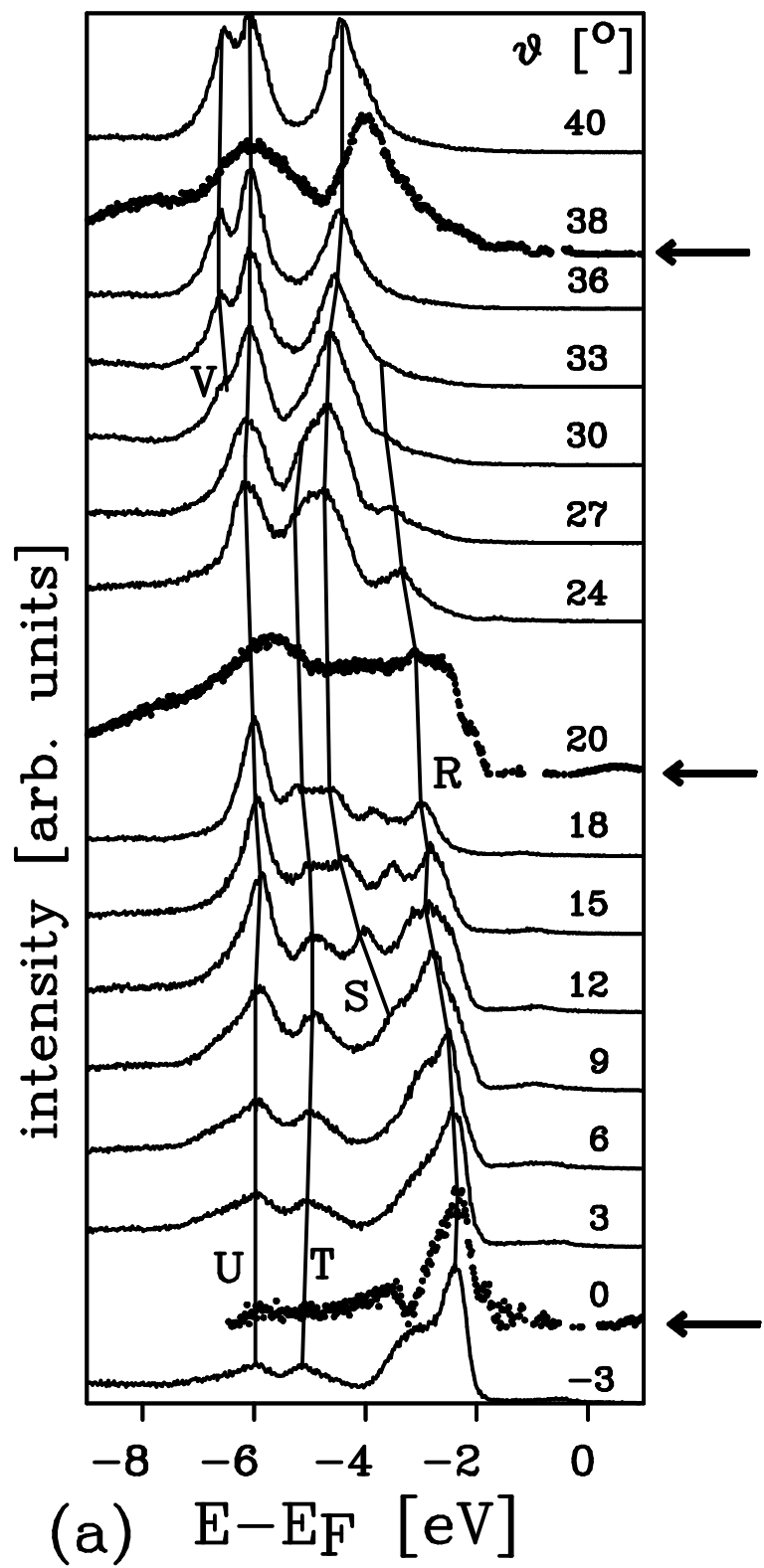
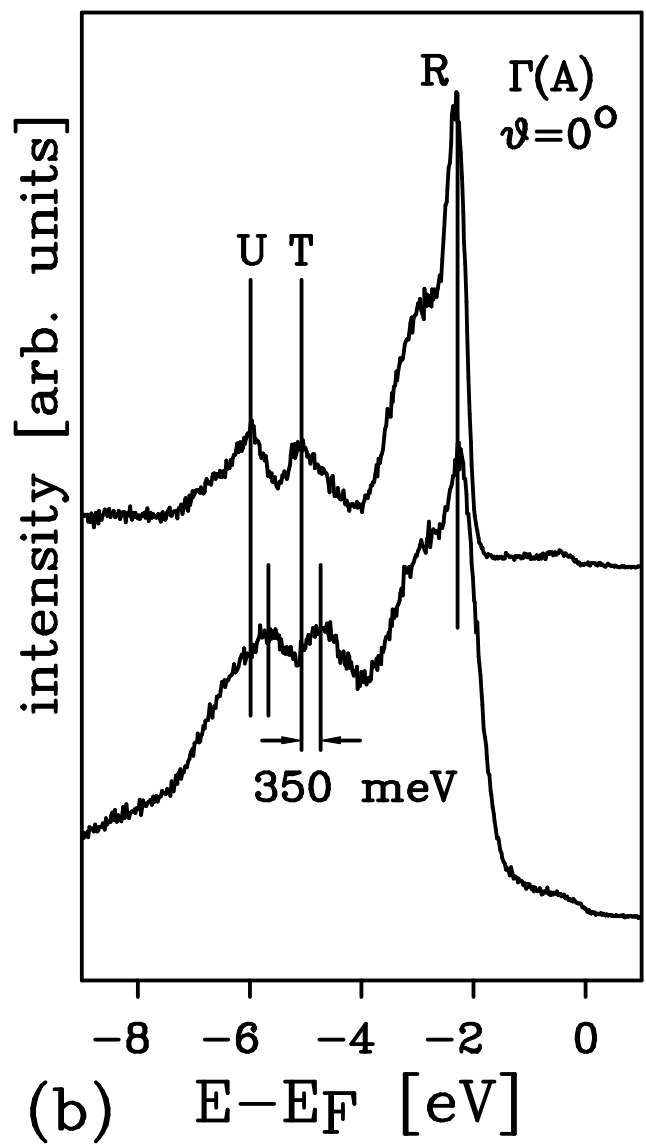
\end{document}